\documentclass[a4paper]{article}
\usepackage{listings}
\usepackage{xcolor}
\usepackage{multirow}
\usepackage{jheppub}  
\usepackage{dcolumn}
\usepackage[english]{babel}
\usepackage[utf8x]{inputenc}
\usepackage[T1]{fontenc}

\usepackage{amsmath}
\usepackage{afterpage}
\usepackage{graphicx, subcaption}
\usepackage[colorinlistoftodos]{todonotes}
\usepackage[colorlinks=true]{hyperref}

\newcommand{\be}{\begin{equation}}  
\newcommand{\ee}{\end{equation}}  
\newcommand{\bea}{\begin{eqnarray}}  
\newcommand{\eea}{\end{eqnarray}}  

\begin{document}

\vspace*{30mm}

\begin{center}
{\LARGE \bf Probing the anomalous $\gamma\gamma\gamma Z$ coupling at the LHC with proton tagging}

\par\vspace*{20mm}\par

{\large \bf C. Baldenegro$^a$, S. Fichet$^b$, G. von Gersdorff$^c$, C. Royon$^a$}

\bigskip

{\em $^a$ University  of Kansas, Lawrence, Kansas, U.S.}
\\
{\em $^b$ ICTP-SAIFR \& IFT-UNESP, R. Dr. Bento Teobaldo Ferraz 271, S\~ao Paulo, Brazil}
\\
{\em $^c$ Departamento de F\'isica, Pontif\'icia Universidade
Cat\'olica de Rio de Janeiro, Rio de Janeiro, Brazil }
\vspace*{5mm}

{\em E-mail: 
c.baldenegro@cern.ch\\
sylvain.fichet@lpsc.in2p3.fr\\
gersdorff@gmail.com\\
christophe.royon@cern.ch}

\vspace*{15mm}

{  \bf  Abstract }

\end{center}
\vspace*{1mm}

\noindent
\begin{abstract}

 The  sensitivities to the anomalous quartic gauge boson coupling $\gamma\gamma\gamma Z$  are estimated via $\gamma Z$ production with intact protons in the forward region at the LHC. Proton tagging proves to be a powerful tool to suppress the background, which allows consideration of the hadronic decays of the $Z$ boson in addition to the leptonic ones.
We discuss the discovery potential for an integrated luminosity  of $300\,\mathrm{fb}^{-1}$ and $3000\,\mathrm{fb}^{-1}$.
The sensitivity we obtain at $300$~fb$^{-1}$ goes beyond the one expected from LHC bounds on the  $Z\rightarrow \gamma\gamma\gamma$ decay  by about three orders of magnitude. 
The $\gamma Z$ channel provides important discriminatory information with respect to the exclusive $\gamma\gamma$ channel,  as many particles beyond the Standard Model (such as a radion or Kaluza Klein gravitons) predict a signal in the latter but not the former.

\end{abstract}

\noindent
\newpage

\section{Introduction}\label{sec:intro}

The Large Hadron Collider currently performs collisions at the unprecedented center-of-mass energy of 13~TeV,  looking for  a signal of  physics beyond the Standard Model (SM). One route to such discovery is to study with high precision the interactions of the Standard Model, by looking for small deviations in the SM couplings. 
Such precision measurements, for instance in the sector of fermion-gauge interactions or more recently of Higgs interactions, have been a focus of intense activity and have proven powerful to constrain New Physics.
 In this work, our interest  lies in the sector of pure gauge interactions. 
    Our focus is the $\gamma\gamma\gamma Z$ interaction, that we study through $\gamma Z$ production via photon fusion with intact outgoing protons in the forward region. Similar photon-induced processes with intact protons have been studied in the past for $W^{+}W^{-}$, $\gamma\gamma$, and other final states in Refs. \cite{usww, usw, Gupta:2011be,Fichet:2013ola, Senol:2013ym, Fichet:2013gsa, Sun:2014qoa, Sun:2014qba,Sun:2014ppa, Senol:2013lca, Sahin:2014dua, Inan:2014mua,Fichet:2015nia,Fichet2015,  Senol:2014vta, Fichet:2016clq,Fichet:2016pvq,Esmaili:2016enf,Khoze:2017igg}.

The measurement we study crucially relies on the 
forward proton detectors in the ATLAS and CMS-TOTEM experiments \cite{ATLAS_detector_paper, Royon:2015tfa}.
We will work in the effective Lagrangian formalism, and the anomalous $\gamma \gamma \rightarrow \gamma Z$ cross section induced by effective operators is given in Sec.~\ref{se:XS}. The contributions to the effective $\gamma \gamma \gamma Z$ interaction from loops of charged particles and tree-level exchange of neutral particles are computed in Sec.~\ref{se:NP}. Turning to $pp$ collisions, a brief description of photon-induced processes  and their measurement in the exclusive channel is discussed in Sec.~\ref{considerations}, and the event generation setup is described in Sec.~\ref{se:generation}. Our analysis in $Z$ hadronic and leptonic channels and the subsequent expected sensitivities are then given in Sec.~\ref{se:selection} and \ref{se:sensitivities}.

\section{ The $\gamma\gamma\gamma Z$ interactions in the SM and beyond}
\label{se:XS}

The $\gamma\gamma\gamma Z$ interaction is induced at one-loop level in the SM via loops of fermions and $W$ bosons.   The amplitudes for the SM $\gamma\gamma \rightarrow \gamma Z$ process have been first computed in Ref.~\cite{Jikia:1993pg}. This process will be neglected in our study as, just like for SM light-by-light scattering, it is greatly reduced in the acceptance of the forward detectors (see sections~\ref{se:proton_det}, \ref{se:exc_bg}).  
The  rare SM decay $Z \rightarrow \gamma\gamma\gamma$ is another process sensitive to the anomalous $\gamma \gamma \gamma Z$ interaction.  It has a branching ratio predicted to be ${\cal B}^{\rm SM}(Z\rightarrow \gamma\gamma\gamma)=5.4 \cdot 10^{-10} $.  The fermion loops have beeen computed in Refs.~\cite{Laursen:1980ba, vanderBij:1988ac} and the $W$ loop contribution in \cite{Baillargeon:1991ia}, the latter is found to be subdominant.

In the presence of New Physics with a mass scale $\Lambda$ heavier than the experimentally accessible energy $E$, all New Physics manifestations can be described using an effective Lagrangian valid for $\Lambda \gg E$. In this low-energy effective field theory (EFT), the $\gamma\gamma\gamma Z$ interactions are described by two  dimension-eight $\gamma\gamma\gamma Z$ operators
\begin{equation}\label{eqn:Lagrangian}
\mathcal{L}_{\gamma\gamma\gamma Z}= \zeta {\cal O}^{\gamma Z} +  \tilde \zeta \tilde {\cal O}^{\gamma Z} =\zeta F^{\mu\nu}F_{\mu\nu}F^{\rho\sigma}Z_{\rho\sigma}+\tilde \zeta F^{\mu\nu} 
\tilde{F}_{\mu\nu}F^{\rho\sigma}\tilde{Z}_{\rho\sigma}\,,
\end{equation}
with $\tilde F^{\mu\nu}=\frac{1}{2}\epsilon^{\mu\nu \rho \sigma} F_{\rho\sigma}$. The ${\cal O}^{\gamma Z}_2=F^{\mu\nu}F_{\nu\rho}F^{\rho\sigma}Z_{\sigma\mu}$  operator can sometimes be encountered in the literature. It is related to the above basis via the identity
$ 4 {\cal O}^{\gamma Z}_2= 2 {\cal O}^{\gamma Z} +\tilde {\cal O}^{\gamma Z} $. The ${\cal O}^{\gamma Z}, \tilde{\cal O}^{\gamma Z}$ provides a somewhat clearer mapping onto the properties of the underlying physics.
   
These operators can be seen as arising from a $SU(2)\times U(1)_Y$ effective Lagrangian with operators such as $B^{\mu\nu}B_{\mu\nu}B^{\rho\sigma}B_{\rho\sigma}$, where $B$ denotes the hypercharge gauge field. The $SU(2)_L\times U(1)_Y$ effective Lagrangian contains ten such operators, see \textit{e.g.} \cite{Belanger:1999aw, Eboli:2006wa, Gupta:2011be, Fichet:2013ola}.
The coefficients of these operators, once expressed with the same Lorentz structures as shown in Eq.~\eqref{eqn:Lagrangian}, must be positive to avoid superluminal excitations in the theory \cite{Adams:2006sv}. 
 The $SU(2)\times U(1)_Y$ effective Lagrangian also generates $4\gamma$, $\gamma \gamma ZZ$ interactions, as described in \cite{Fichet:2013ola}. Because of the large number of effective operators in the $SU(2)\times U(1)_Y$ Lagrangian, anomalous interactions in the broken phase can be considered as independent.

\begin{figure}
\centering
\includegraphics[width=0.3\textwidth]{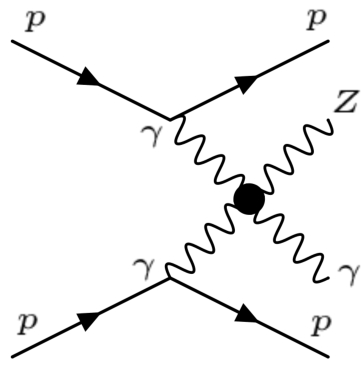}
\caption{\label{AAAZ_feynman} Anomalous $\gamma Z$ production via photon fusion with intact protons in the final state.}
\end{figure}

The  operators of Eq.~\eqref{eqn:Lagrangian} induce an anomalous $Z\rightarrow \gamma\gamma\gamma$ decay \cite{Baillargeon:1995dg}, with a partial width that in our notation reads
\be
{\Gamma}^{\rm NP}(Z\rightarrow \gamma\gamma\gamma)=\frac{m_Z^9 (2\zeta^2+2\tilde\zeta^2- \zeta \tilde \zeta)}{8640 \pi^3} \,.
\ee
An anomalous $\gamma\gamma\rightarrow \gamma Z$ reaction is also induced, which is the focus of this work. 
We find the unpolarized differential cross section  to be
 \footnote{It has been noted in \cite{Baillargeon:1995dg} that the operators ${\cal O}_\pm={\cal O}^{\gamma Z}\pm \tilde{\cal O}^{\gamma Z}$   do not interfere. This property provides a cross check of our result Eq.~\eqref{eqn:cross_section}, as in this basis we get $\zeta_\pm=\zeta\pm \tilde \zeta$,  $(3\zeta^2+3\tilde \zeta^2-2\zeta\tilde \zeta)=\zeta^2_++ 2\zeta_-^2 $ and 
   $4(\zeta^2+\tilde \zeta^2-\zeta\tilde\zeta)=\zeta^2_++ 3\zeta_-^2$ 
   , hence a vanishing interference.  }
\be
\frac{\mathrm{d}\sigma^{\rm NP}_{\gamma\gamma\rightarrow \gamma Z}}{\mathrm{d}\Omega} = \frac{\beta}{ 16 \pi^2 s} \Big[ (3\zeta^2+3\tilde \zeta^2-2\zeta\tilde\zeta)(st +tu+us)^2
-4(\zeta^2+\tilde \zeta^2-\zeta\tilde\zeta)^2m_Z^2stu\Big]\,,
\label{eqn:cross_section}
\ee
where $s$, $t$, and $u$ are the usual Mandelstam variables and $\beta = 1-m_Z^2/s$ for the $\gamma Z$ final state.

As the EFT is nonrenormalizable, a breakdown of unitarity is expected at high energies. Using the well-known partial wave analysis \cite{Agashe:2014kda} we can estimate for what values of $\zeta$, $\tilde \zeta$ and $s$ the theory remains unitary. By imposing unitarity on the $S$-wave of the EFT  amplitudes and neglecting the $Z$ boson mass one finds the conditions (see \cite{Fichet:2013ola} for details on similar amplitudes)
\bea
|\zeta+\tilde \zeta|s^2<4\pi\,,\quad |\zeta-\tilde \zeta|s^2<\frac{12 \pi}{5}\,.  
\eea
 As most of the recorded $\gamma Z$ events have $\sqrt{s}$ below 1 TeV, we expect the EFT to remain unitary for couplings up to 
\be
\zeta, \tilde \zeta < (10^{-12}-10^{-11}) {\rm\ GeV}^{-4}\,.
\label{unitarity}
\ee
The sensitivities we will derive in Sec.~\ref{se:sensitivities} are much lower than these unitarity bounds. However, as a caveat, we stress that unless the underlying New Physics model is very strongly coupled, the EFT typically breaks down before unitarity is violated.

\section{ Contributions from New Physics}
\label{se:NP}

Loops of heavy particles charged under $SU(2)_L\times U(1)_Y$ contribute to the $\gamma\gamma\gamma Z $ couplings.
These loop contributions only depend on the mass and quantum numbers of the particle in the loop and can thus be given in full generality. Denoting hypercharge by $Y$, sine and cosine of the Weinberg angle by $s_w$ and $c_w$ and labeling the $SU(2)_L$ representation by its dimension $d$,  we can write  \cite{Fichet:2013ola}
\be \label{eq:loop}
\Big(\zeta, \tilde \zeta\Big) = \Big(c_{s}, \tilde c_{s}\Big)\, \frac{ \alpha^2_{\rm em} }{ s_w c_w \, m^4 } \, d \left( c_w^2 \frac{3d^4-10d^2+7}{240} +  (c_w^2-s_w^2)\frac{(d^2-1)Y^2}{4} - s^2_w Y^4
  \right) \,, 
\ee
with
\be
c_{s}=
\begin{cases}
\frac{7}{360} & s=0 \\
\frac{2}{45} & s=\frac{1}{2} \\
\frac{29}{40} & s=1 \\
\end{cases}
\,,\quad
\tilde c_{s}=
\begin{cases}
\frac{1}{360} & s=0 \\
\frac{7}{90} & s=\frac{1}{2} \\
\frac{27}{40} & s=1 \\
\end{cases}
\ee
where $s$ denotes the spin of the heavy particle running in the loop. \footnote{
The coefficients $c_s$, $\tilde c_s$ have been determined Ref.~\cite{Baillargeon:1995dg} in a specific case. }

Beyond perturbative contributions to $\zeta$, $\tilde \zeta$ from  
charged particles,  non-renormalizable interactions of neutral particles are 
also present in common extensions of the SM.  Such theories can contain 
scalar, pseudo-scalar and spin-2 resonances, respectively denoted $\varphi$, 
$\tilde \varphi$, $h^{\mu\nu}$, that couple  to the photon 
as
\begin{equation} \begin{split} 
\mathcal L_{\gamma\gamma}= &\varphi\, 
\left[\frac{1}{f_{0^+}^{\gamma\gamma}}\, (F_{\mu\nu})^2+\frac{1}{f_{0^+}^{\gamma Z}}\, F_{\mu\nu} Z_{\mu\nu}\right]
+ \tilde\varphi  \, \left[ \frac{1}{f_{0^-}^{\gamma\gamma}}\, F_{\mu\nu} \tilde F_{\mu\nu}\,
+\frac{1}{f_{0^-}^{\gamma Z}}\, F_{\mu\nu} \tilde Z_{\mu\nu}\,
 \right] 
\\&  +  h^{\mu\nu}\, \left[ \frac{1}{f_{2}^{\gamma\gamma}} (-F_{\mu\rho} 
F_{\nu}^{\,\,\rho}+\eta_{\mu\nu} (F_{\rho\lambda})^2/4) +
\frac{1}{f_{2}^{\gamma Z}} (-F_{\mu\rho} 
Z_{\nu}^{\,\,\rho}+\eta_{\mu\nu} F_{\rho\lambda}Z_{\rho\lambda}/4)\right]\,,
\end{split}
\end{equation}
and generate  the $\gamma\gamma\gamma Z$ couplings by tree-level exchange as 
\be (\zeta, \tilde \zeta)= \frac{1}{f^{\gamma\gamma}_{s}f^{\gamma Z}_{s}\, m^2} (d_{ s}, \tilde d_{ s})  \ee where
\be
d_{s}=
\begin{cases}
1 & s=0^+ \\
 0 & s=0^- \\
\frac{1}{4} & s=2\\
\end{cases}
\,,\quad
\tilde d_{s}=
\begin{cases}
0 & s=0^+ \\
1  & s=0^- \\
\frac{1}{4} & s=2 \\
\end{cases}
\,.
\ee

Interestingly, the $f^{\gamma Z}$ couplings vanish if the neutral particle  couples universally   to  the $W^I$ and $B$ kinetic terms of the $SU(2)_L\times U(1)_Y$ Lagrangian. This happens in particular when the neutral particle couples to gauge bosons via the stress-energy tensor.  It is the case of the 
 Kaluza Klein (KK) graviton present in models of warped extra dimensions with gauge fields on the IR brane, as well as the radion and the KK graviton in  bulk gauge field scenarios with small IR brane kinetic terms \cite{Fichet:2013ola,Fichet:2013gsa}.
 
This peculiar feature of the $\gamma \gamma \gamma Z$  coupling becomes very interesting when put together with the measurement of the $4\gamma$ interaction. Indeed, if a $\gamma \gamma$ signal was observed, the $\gamma Z$ channel would then  provide a clear test  whether or not the underlying exchanged particle is universally coupled to the gauge kinetic terms.

\section{The $pp \rightarrow \gamma Z\, pp $ process at the LHC}\label{considerations}
\subsection{Photon coherent flux}

We use the equivalent photon approximation  \cite{Terazawa:1973tb}, \cite{Budnev} to describe the $pp \rightarrow \gamma Z\, pp $ process via photon exchanges. In this approximation, the almost real photons (with low virtuality $Q^2=-q^2$) are emitted by the incoming protons producing a state $X$, $pp\rightarrow pXp$, through photon fusion $\gamma\gamma\rightarrow X$. The photon spectrum of virtuality $Q^2$ and energy $E_\gamma$ is proportional to the fine-structure constant $\alpha_{\rm em}$ and reads:
\begin{equation}
dN = \frac{\alpha_{\rm em}}{\pi} \frac{\mathrm{d}E_\gamma}{E_{\gamma}} \frac{\mathrm{d}Q^2}{Q^2}\bigg[ \bigg(1 - \frac{E_\gamma}{E}\bigg) \bigg(1-\frac{Q^2_{\mathrm{min}}}{Q^2}\bigg )F_E + \frac{E_\gamma^2}{2E^2} F_M \bigg  ]
\end{equation}
where $E$ is the energy of the incoming proton of mass $m_p$, $Q^2_{\mathrm{min}}=m_p^2E_\gamma^2/[E(E-E_\gamma)]$ the photon minimum virtuality allowed by kinematics and $F_E$ and $F_M$ are functions of the electric and magnetic form factors $G_E$ and $G_M$. In the dipole approximation the latter read
\begin{equation}
F_M=G^2_M  \qquad F_E=(4m_p^2G^2_E+Q^2G^2_M)/(4m_p^2+Q^2)\qquad G^2_E=G^2_M/\mu_p^2=(1+Q^2/Q^2_0)^{-4} \,.
\label{eq:newera:elmagform}
\end{equation}
The magnetic moment of the proton is $\mu_p^2=7.78$ and the fitted scale $Q^2_0=0.71$ GeV$^2$.
Since the electromagnetic form factors fall steeply as a function of $Q^2$,
the cross section can be factorized into the matrix element of the photon fusion process  and 
the two photon fluxes. In order to obtain the production cross section, the photon fluxes are first integrated over $Q^2$
\begin{equation}
f(E_\gamma)=\int^{\infty}_{Q^2_{\rm min}}\frac{d N}{d E_\gamma d Q^2} d Q^2 \,.
\label{sm:flux_q2}
\end{equation}
 The result is given for instance in Ref.~\cite{usww}. These fluxes can then be used to introduce an effective differential luminosity $d L^{\gamma\gamma} \slash d W$ obtained by integrating $f(E_{\gamma1})\,f(E_{\gamma2})\,d E_{\gamma1}\,dE_{\gamma2}\, \delta(W-2\sqrt{E_{\gamma1} E_{\gamma2}})$ where $W$ is  invariant mass of the diphoton system.
Using the effective photon
luminosity, the total cross section for the $pp\rightarrow pXp $ process reads
\begin{equation}
\sigma=\int\sigma_{\gamma\gamma\rightarrow X}
      \frac{\mathrm{d} L^{\gamma\gamma}}{\mathrm{d} W}d W 
\label{eq:sm:totcross}
\end{equation}
where $\sigma_{\gamma\gamma\rightarrow X}$ denotes the 
cross section of the sub-process $\gamma\gamma\rightarrow X$, dependent on the 
invariant mass of the two-photon system.
In addition to the photon exchange, there might be additional soft gluon exchanges that might destroy the protons. To take into account this effect, we can introduce the so-called survival probability that the protons remain intact in photon-induced processes \cite{survival1}, \cite{survival2}. In this paper, we assumed a survival probability of 90\%.

\subsection{Proton detectors} \label{se:proton_det}

The $pp \rightarrow p \gamma Z p$ proccess can be probed via the detection of two intact protons in the forward proton detectors at CT-PPS or AFP \cite{ATLAS_detector_paper}, \cite{cms} and the detection of the $Z$ boson decay and the photon in the respective central detector. The forward detectors are located symmetrically at about 210 m from the main interaction vertex and cover a range of $0.015 < \xi_{1,2} < 0.15$, where $\xi_{1,2}$ is the proton fractional momentum loss which is reconstructed by the forward proton detector. 
This leads to an acceptance in the central mass $m_{\gamma Z}$ between $300$ and $1900$ GeV.

The average number of multiple proton-proton collisions per bunch crossing, $\mu$, sets a huge background environment on the search for exclusive events. Intact forward protons arising from the pile-up together with an uncorrelated non-exclusive process in the central detector can mimic the signal exclusive events. Usually $\mu$ ranges from 30 to 50 interactions per bunch crossing at the current LHC luminosity. The pile-up is expected to go up to $\mu=200$ interactions at the High Luminosity LHC which will pose a challenge on the search for New Physics. The detection and characterization of the outgoing protons  provides the complete kinematic information on the event, which in turn allows us to exploit the four-momentum conservation  
via rapidity and mass matching methods. These methods provide a strong background rejection which is the key feature of the forward proton detectors in exclusive processes. Further background rejection can be achieved with the use of timing detectors. Timing detectors are expected to be installed and operating in both  CT-PPS and ATLAS to measure the time-of-flight of protons with a precision of $\sim$ 15 ps, which would allow to determine the interaction vertex of the protons with a 2.1 mm precision, thus allowing a large background rejection by a factor of $\sim 40$ \cite{Timing}. In this work, we will not make use of this potential future improvement.

\subsection{Central detection of $Z\gamma$}

In ATLAS and CMS, photons can be reconstructed in the central detectors instrumented with electromagnetic calorimeters which cover the pseudorapidity range $|\eta| < 2.5$ and provide excellent resolution in terms of energy (less than a percent at $p_T > 100$ GeV) and position (0.001 units of $\eta$ and 1 mrad on the azimuthal angle $\phi$) for photons and leptons with $p_T$ ranging from a few GeV up to the TeV scale. Photon identification efficiency is expected to be around 75\% for $p_T > 100$ GeV. In addition, about 1\% of the electrons and jets are misidentified as photons \cite{Khachatryan:2015iwa}.

The decay of the Z boson  is widely studied in CMS and ATLAS and is used for calibration of the detectors \cite{Barter:2016oan}.
In this study we consider both leptonic (electrons and muons)  and hadronic decays.  For ATLAS, the fiducial acceptance corresponds to leptons with tranverse momenta $p_T^\ell>25$ GeV and absolute rapidity $\eta_\ell < 2.5$. For Z boson production, the dilepton invariant mass $m_{\ell\bar{\ell}}$ is required to be between $66 < m_{\ell\bar{\ell}} < 116$ GeV. Similar requirements are made in CMS. 
For the $Z$ boson decay into hadrons, the jet is reconstructed by clustering particles deposited in the electromagnetic and hadronic calorimeters. The energy of photons is obtained directly from the electromagnetic calorimeter measurement. The energy of a charged hadron is determined from a combination of the track momentum and the corresponding electromagnetic and hadronic calorimeter energies. The energy of a neutral hadron is obtained from the calibrated energies in electromagnetic and hadronic calorimeters. The typical jet energy resolution is between 5-10\% for jets with $p_T>200$ GeV. Commonly, the anti-$k_T$ jet clustering-algorithm with a radius parameter of $R=0.5$ is used in CMS and ATLAS \cite{Schroeder:2015ila,Berta:2016ukt}.

\section{Event generation} \label{se:generation}

The anomalous $\gamma\gamma\rightarrow \gamma Z$ process with intact protons in the final state has been implemented in the Forward Physics Monte Carlo (FPMC) generator \cite{FPMC}. Contributions and simulations of the various backgrounds are discussed in the subsections below. 
In order to model systematic uncertainties on final states, we have included Gaussian smearings on the photon, lepton and hadron energies of 1\%. In addition, we apply Gaussian smearings of 15 \% in the individual reconstructed jet energy, as well as 0.1\% for the pseudorapidity and 1 mrad for the azimuthal angle.

\subsection{Pile-up backgrounds }

The largest background to the $p\gamma Z p $ final state originates from $\gamma Z$ detection in the central detectors simultaneously with the detection of two intact protons from pile-up.

Background contribution in the $jj \gamma$ channel is dominated by $W^{\pm}\gamma$ and $\gamma Z$ in association with pile-up protons if we restrict ourselves to two-jet final states. Around 1\% of the electrons are misidentified as photons, thus the background $q\bar{q}e$ in association with pile-up is also considered in our study. The non-exclusive background processes were simulated in PYTHIA8 \cite{Sjostrand:2014zea} at leading order. Jets are reconstructed with the anti-$k_t$ clustering-algorithm with FastJet \cite{Fastjet} using $R=0.5$ and $p_{T min}=10$ GeV at the hadron level, which is close to the standard CMS and ATLAS parameter choice for jet reconstruction. We assign a 15\% resolution to the reconstructed jets energy, and apply smearings on the $\phi_j$, $p_{T j}$ and $\eta_j$ on top of the gaussian smearings applied to the individual particles that form the jet.

In the $\ell\bar{\ell}\gamma$ channel, the dominant background is the leptonic decay of the non-exclusive $\gamma Z$ production in association with pile-up protons, as can be seen in Figure \ref{central_mass_leptons}. We also consider misidentification of jets and electrons as photons as part of the background.

The pile-up events were simulated as follows. For each non-exclusive background event generated on PYTHIA, the number of pile-up interactions in the event is drawn from a Poisson distribution with mean $\mu=50,200$ respectively for the low and high luminosity scenarios.  We draw the tag probability for the protons arising from the pile-up interactions from a uniform distribution and compare it with the single, double or no tag probabilities, which were computed by propagating single and double diffractive protons generated on PYTHIA8 along the beamline up to the proton detectors within their acceptance (see \cite{Kepka_pileup, Timing} for more details). We assign the fractional momentum loss $\xi=\Delta p/p$, which would be reconstructed by the forward detectors in AFP or TOTEM, by randomly sampling the distribution $f(\xi) = 1/\xi$ (Which roughly resembles the $\xi$ distribution measured by the proton detectors) defined in $0.015<\xi<0.15$ via its inverse transform. When the forward detectors have at least one proton tagged in each arm, we compare the diproton mass $\sqrt{\xi_1\xi_2 s}$ and rapidity $\frac{1}{2}\log(\xi_1/\xi_2)$ with the central mass and rapidity of the $\gamma Z$ final state, and select the best match in such observables. Only events with two proton tags pass through the selection cuts quoted in Tables \ref{qqbar_table}, \ref{llbar_table} and \ref{llbar_table_highlumi}.

\subsection{Exclusive backgrounds} \label{se:exc_bg}

The SM predicts  $Z\gamma$ production at one-loop with two intact protons via two mechanisms: elastic gluon fusion \cite{Khoze:2001xm}, with the exchange of an extra gluon to ensure the process is colourless and the protons stay intact, and photon fusion.

The elastic gluon fusion contribution can be neglected at high  $m_{Z\gamma}$ (within the proton taggers acceptance) since the soft gluon emission in the gluon ladder has to be suppressed in order to get an exclusive diffractive event with intact protons. In practice, a Sudakov form factor is introduced to suppress this emission which greatly reduces the cross section at high  $m_{Z\gamma}$ \cite{Khoze:2001xm}. This is true in general for central exclusive processes in QCD.

The SM cross section for $\gamma\gamma \rightarrow Z\gamma$ is related to the $\gamma \gamma$ cross section roughly as $\sigma_{\gamma\gamma\rightarrow Z\gamma}^{\rm SM} \approx 6 \sigma_{\gamma\gamma\rightarrow \gamma\gamma}^{\rm SM}$ within the mass acceptance (See for instance Figures 4 and 5 in \cite{Jikia:1997yt} ). We can use this to extrapolate the results from the diphoton studies in Ref. ~\cite{Fichet2015} (From Table 2, last row). The expected number of events for exclusive $Z\gamma$ in the $jj\gamma$ ($\ell\bar{\ell}\gamma$) channel is $\sim$ 0.25 ($\sim$ 0.04) for an integrated luminosity of 300 fb$^{-1}$ after selection cuts, which is found to be less than 10\% of the total background in the exclusive channel as seen in Sec. \ref{se:selection}. For 3000 fb$^{-1}$, we expect less than $0.25$ background events in $\ell\bar{\ell}\gamma$, less than 12\% of the total background estimated in Sec. \ref{se:selection}. Thus, in this study we assume that the QCD and QED exclusive $Z\gamma$ contribution is small and focus on the non-exclusive background contribution, which constitutes the dominant background in this search.

\section{Event selection} \label{se:selection}

\begin{figure}[]
\centering
\includegraphics[width=0.7\textwidth]{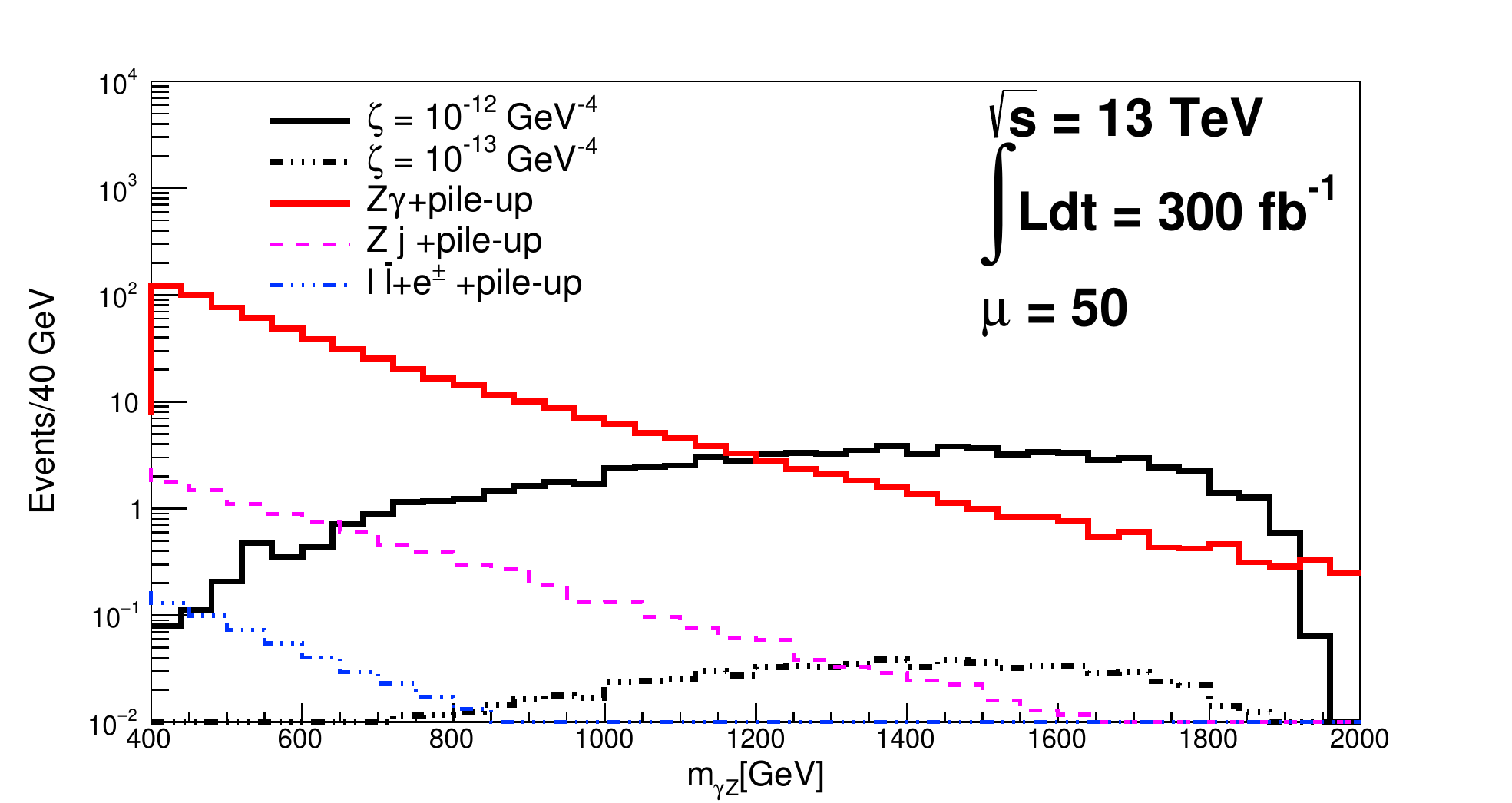}
\caption{ \label{central_mass_leptons} $\gamma Z$ mass distribution for the signal in the $\ell\bar{\ell}\gamma$ channel for two coupling values ($\zeta  = 10^{-12}, 10^{-13}$ GeV$^{-4}$) for events within the $0.015<\xi<0.15$ proton detectors acceptance and the requirement on transverse momenta $p_{T \gamma},p_{T \ell\bar{\ell}}>100$ GeV. The main contribution to the background is the SM $Z\gamma$ production in association with protons arising from the pile-up. The plot assumes an integrated luminosity of $300\,\mathrm{fb}^{-1}$ and  an average pile-up of $\mu=50$.}
\end{figure}

The  $(jj),\gamma$ $(\ell\bar{\ell}),\gamma$ final states from the New Physics signal are typically back-to-back and have similar transverse momenta. In the dijet (dilepton) case, this translates to cuts on $|\Delta \phi_{jj\gamma} - \pi|<0.02$ ($|\Delta \phi_{\ell\bar{\ell}\gamma} - \pi|<0.02$) and $p_{T \gamma}/p_{T  jj}>0.90$ ($p_{T \gamma}/p_{T \ell\bar{\ell}})>0.95$). As can be seen in Figures \ref{central_mass_leptons} and \ref{central_mass_jets}, the signal events appear in the high mass region, allowing for further background rejection  by asking $m_{jj\gamma}>700$ GeV ($m_{\ell\bar{\ell}\gamma} > 600 $ GeV). 
The probability to detect at least one proton in each of the forward detectors is estimated to be 32\%, 66\% and 93\% for 50, 100 and 200 additional interactions respectively. The pile-up background is further suppressed by requiring the proton missing invariant mass $m_{pp}$ to match the $Z\gamma$ invariant mass within 10\% (5\%) resolution, $m_{\gamma jj} = \sqrt{\xi_1\xi_2s} \pm 10\%$ ($m_{\gamma \ell\bar{\ell}} = \sqrt{\xi_1\xi_2s} \pm 5\%$) and the $\gamma Z$ system rapidity and the rapidity of the two protons $y_{pp} = \frac{1}{2}\ln(\xi_1/\xi_2)$ to be the same within $|y_{\gamma jj}-y_{pp}|<0.10$ ($|y_{\gamma \ell\bar{\ell}}-y_{pp}|<0.03$) units in rapidity for the hadronic (leptonic) channel. This is the key background rejection tool provided by the forward detector information.
The number of expected signal and background events passing their respective selections can be seen in Table \ref{qqbar_table}, (Table \ref{llbar_table}) for the $jj\gamma$ channel ($\ell\bar{\ell}\gamma$ channel) for an integrated luminosity of 300\,fb$^{-1}$ ($\approx 3$ years of data-taking at the LHC) and moderate pile-up interactions $\mu=50$ at $\sqrt{s}=13$ TeV.

\begin{table}[p]
\centering
\begin{tabular}{|c||c|c||c|c|c|}
\hline
\multirow{3}{*}{Cut/Process} & \multirow{3}{*}{Signal} & \multirow{3}{*}{Signal} & \multirow{3}{*}{$\gamma Z$} & \multirow{3}{*}{$W^{\pm} \gamma$} & \multirow{3}{*}{$jje^\pm$} \\
                  &                   &                   &                   &                   &                   \\
                  &            $\zeta$ ($\tilde \zeta=0$)       &      $\zeta = \tilde \zeta$             &       +pile-up            &   +pile-up                &      +pile-up             \\ \hline \hline
\multirow{3}{*}{$0.015<\xi_{1,2}<0.15$, $p_{T \gamma}>150$ GeV} & \multirow{3}{*}{38.6} & \multirow{3}{*}{51.4 } & \multirow{3}{*}{1951.8} & \multirow{3}{*}{1631} & \multirow{3}{*}{8.47}\\
                  &                        &                       &                         &              &              \\
$p_{T jj}>100$ GeV]                   &                          &            &        &      &                          \\ \hline
$m_{\gamma Z}>700$ GeV                  & 37                  & 49.5                 & 349.8                  & 358.9           & 1.3                  \\ \hline
\multirow{3}{*}{$p_{T jj}/p_{T \gamma}>0.90$,} & \multirow{3}{*}{33.8} & \multirow{3}{*}{45.1} & \multirow{3}{*}{144.7}   & \multirow{3}{*}{145.4}   &   \multirow{3}{*}{0.54} \\
                  &                        &                       &                             &                   &          \\
$|\Delta \phi - \pi | < 0.02$                  &                        &                       &                     &                &                    \\ \hline
$\sqrt{\xi_1\xi_2 s}=m_{\gamma Z}\pm 10\%$                  & 28.2                  & 35.7                  &           19.7           & 19.3                & 0.1             \\ \hline
$|y_{pp}-y_{\gamma Z}|<0.05$                  & 25.5                 & 32.7                  &   1.5                     &  1.6              &     0                  \\ \hline
\end{tabular}
\caption{ \label{qqbar_table} Number of signal and background events in the $jj\gamma$ channel after the selection cuts for an integrated luminosity of 300\,fb$^{-1}$ and $\mu=50$ at $\sqrt{s}=13\,$TeV, and for $\zeta  = 4 \cdot 10^{-13} \, \mathrm{GeV}^{-4}$.
Non-exclusive events were simulated on PYTHIA8 at leading-order and signal events in the FPMC. Jets are reconstructed with the anti-$k_t$ clustering-algorithm using $R=0.5$ and $p_{T\,{ \rm min }} = 10$ GeV. }
\end{table}

\begin{table}[p]
\centering
\begin{tabular}{|c||c|c||c|c|c|}
\hline
\multirow{3}{*}{Cut/Process} & \multirow{3}{*}{Signal} & \multirow{3}{*}{Signal} & \multirow{3}{*}{$\gamma Z$} & \multirow{3}{*}{$\ell\bar{\ell}j$} & \multirow{3}{*}{$\ell \bar{\ell} e^\pm$} \\
                  &                   &                   &                   &                   &                   \\
                  &            $\zeta$ ($\tilde \zeta=0$)       &      $\zeta = \tilde \zeta$             &       +pile-up            &   +pile-up                &      +pile-up             \\ \hline \hline
\multirow{3}{*}{[$0.015<\xi_{1,2}<0.15$, $p_{T \gamma}>100$ GeV} & \multirow{3}{*}{13.2 }  & \multirow{3}{*}{17.4} & \multirow{3}{*}{2239.2}  & \multirow{3}{*}{64.5}  & \multirow{3}{*}{1.2} \\
                  &                          &                          &         &          &                 \\
$p_{T \ell\bar{\ell}}>100$ GeV]                   &                          &            &        &      &                          \\ \hline
$m_{\gamma Z}>600$ GeV                  & 12.9                & 17.1                 & 227    & 3.8 & 0.2               \\ \hline
\multirow{3}{*}{$p_{T \gamma}/p_{T \ell\bar{\ell}}>0.95$,} & \multirow{3}{*}{12.6 } & \multirow{3}{*}{16.7} & \multirow{3}{*}{175} & \multirow{3}{*}{0} & \multirow{3}{*}{0} \\
                  &                          &                          &         & &                 \\
$|\Delta \phi - \pi | < 0.02$                  &                          &                          &          &  &                \\ \hline
$\sqrt{\xi_1\xi_2 s}=m_{\gamma Z}\pm 5\%$                  & 12.2              & 16.4               & 12.7  & 0        & 0   \\ \hline
$|y_{pp}-y_{\gamma Z}|<0.03$                  & 10             & 13.7                 & 0.6         & 0  & 0        \\ \hline
\end{tabular}
\caption{ \label{llbar_table} Same as Tab.~\ref{qqbar_table} for the $\ell\bar\ell\gamma$ channel. The selection yields a signal efficiency of about $75\%$ with an essentially background-free measurement in this channel.}
\end{table}

\begin{table}[p]
\centering
\begin{tabular}{|c||c|c||c|c|c|}
\hline
\multirow{3}{*}{Cut/Process} & \multirow{3}{*}{Signal} & \multirow{3}{*}{Signal} & \multirow{3}{*}{$\gamma Z$} & \multirow{3}{*}{$\ell\bar{\ell}j$} & \multirow{3}{*}{$\ell \bar{\ell} e^\pm$} \\
                  &                   &                   &                   &                   &                   \\
                  &            $\zeta$ ($\tilde \zeta=0$)       &      $\zeta = \tilde \zeta$             &       +pile-up            &   +pile-up                &      +pile-up             \\ \hline \hline
\multirow{3}{*}{[$0.015<\xi_{1,2}<0.15$, $p_{T \gamma}>200$ GeV} & \multirow{3}{*}{99.5}  & \multirow{3}{*}{132.8} & \multirow{3}{*}{6403.6} & \multirow{3}{*}{1207.3} & \multirow{3}{*}{30.1} \\
                  &                          &                          &             &         &                     \\
$p_{T \ell\bar{\ell}}>200$ GeV]                  &                          &            &        &      &                          \\ \hline
$m_{\gamma Z}>1100$ GeV                  & 77.2                 & 105.5                  & 550.4   &    106.3  & 5.3                \\ \hline
\multirow{3}{*}{$p_{T \gamma}/p_{T \ell\bar{\ell}}>0.95$,} & \multirow{3}{*}{76.4 } & \multirow{3}{*}{104.8 } & \multirow{3}{*}{458.2} & \multirow{3}{*}{19.2} & \multirow{3}{*}{0.5} \\
                  &                          &                          &         &       &                 \\
$|\Delta \phi - \pi | < 0.01$                  &                          &                &        &       &                          \\ \hline
$\sqrt{\xi_1\xi_2 s}=m_{\gamma Z}\pm 3\%$                  & 62.2                 & 85                  & 16.4         & 1.2        & 0                 \\ \hline
$|y_{pp}-y_{\gamma Z}|<0.025$                  & 46.2               & 72                  & 1.8        & 0         &     0           \\ \hline
\end{tabular}
\caption{ \label{llbar_table_highlumi} 
Same as Tab.~\ref{llbar_table}, but with an integrated luminosity of 3000 fb$^{-1}$.
We are left with $\approx 2$ background events after all the selection cuts with a signal efficiency of $50\%$.
}
\end{table}

\clearpage


\begin{figure}
\centering
\includegraphics[scale=0.51]{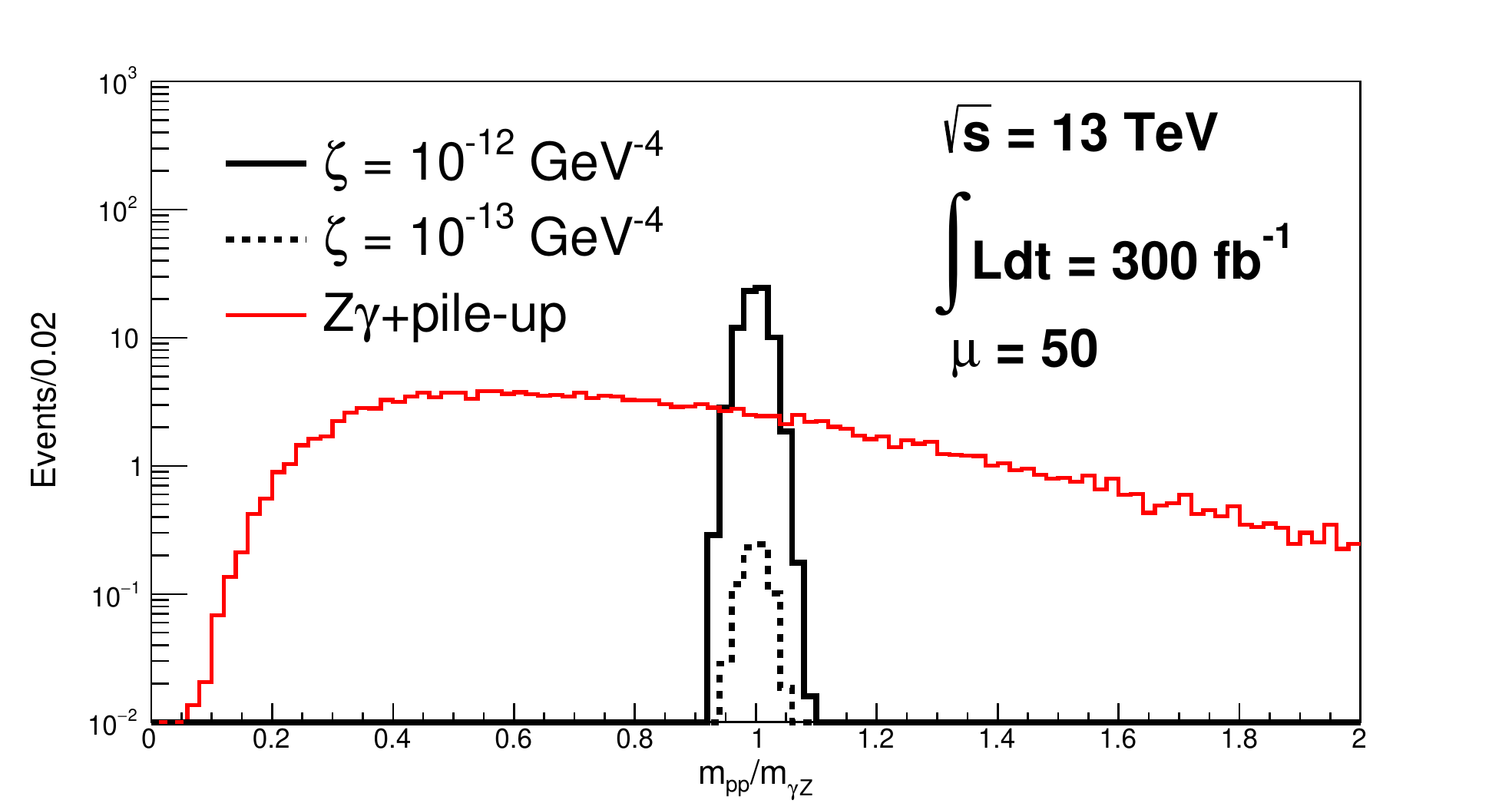}
\includegraphics[scale=0.51]{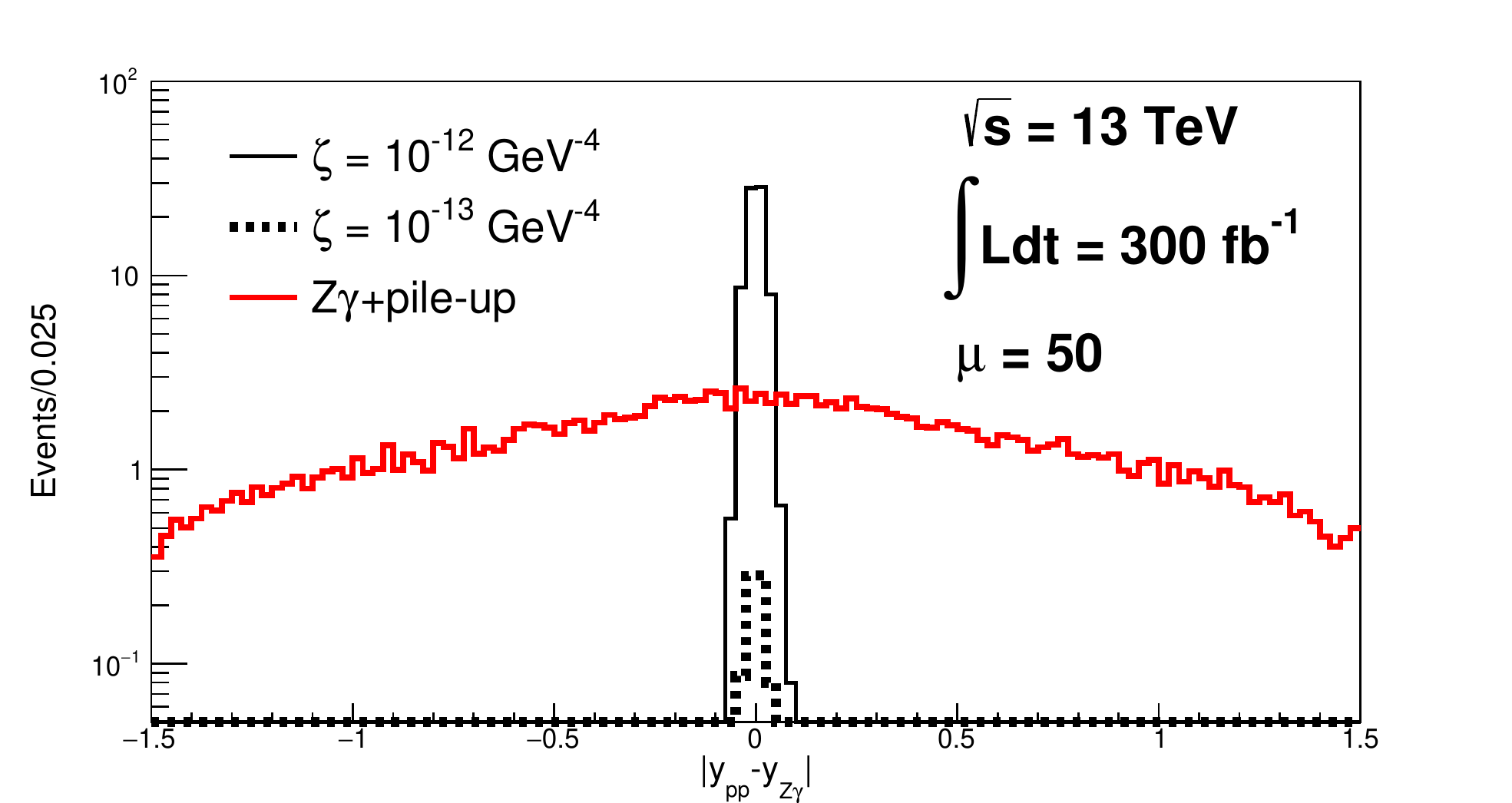}
\caption{\label{exclusivity_leptons} Missing diproton mass $m_{pp} = \sqrt{\xi_1\xi_2 s}$ to central mass ratio distribution (Top) and rapidity difference distribution (Bottom) in the $\ell\bar{\ell}\gamma$ channel for signal and background within the acceptance $0.015<\xi_{1,2}<0.15$ considering two different coupling values after applying the requirements on the acceptance, $p_T$, invariant mass $m_{\gamma Z}$, $p_T$ ratios and angle separation according to Table \ref{llbar_table}. The width of the signal is due mainly to the $\xi_{1,2}$ resolution. The integrated luminosity is $300$ fb$^{-1}$   and the average pile-up  is $\mu = 50$.}
\end{figure}

\clearpage

\begin{figure}
\centering
\includegraphics[width=0.65\textwidth]{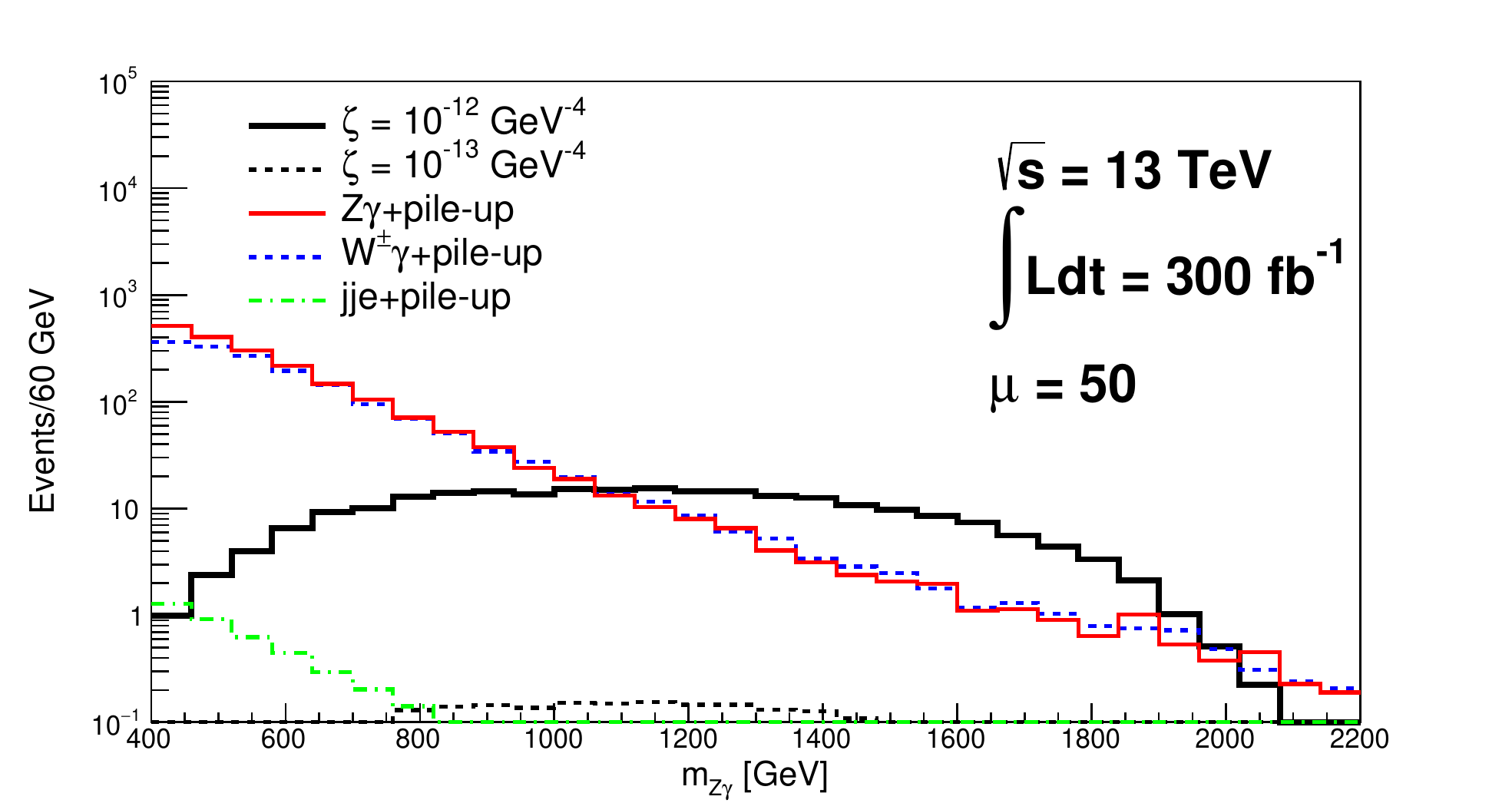}
\caption{ \label{central_mass_jets} $\gamma Z$ mass distribution for the signal in the $jj\gamma$ channel for two coupling values ($\zeta  = 10^{-12}, 10^{-13}$ GeV$^{-4}$, $\tilde\zeta =0$) for events within the $0.015<\xi_{1,2}<0.15$ proton detectors acceptance and after the transverse momenta requirement as in Table \ref{qqbar_table}. The plot assumes an integrated luminosity of 300 fb$^{-1}$ and an average pile-up  of $\mu = 50$.}
\end{figure}

\begin{figure}
\centering
\includegraphics[scale=0.45]{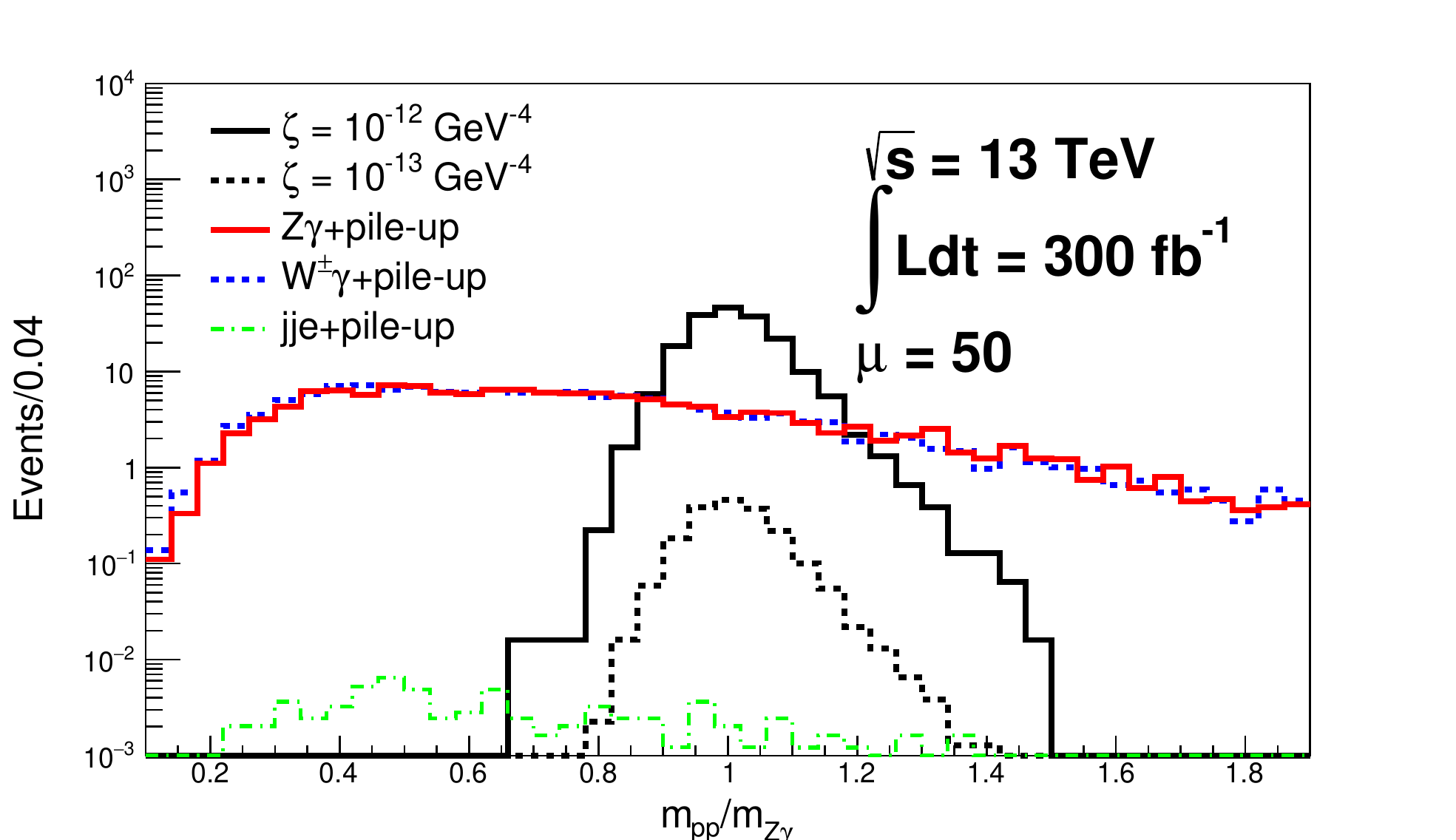}
\includegraphics[scale=0.45]{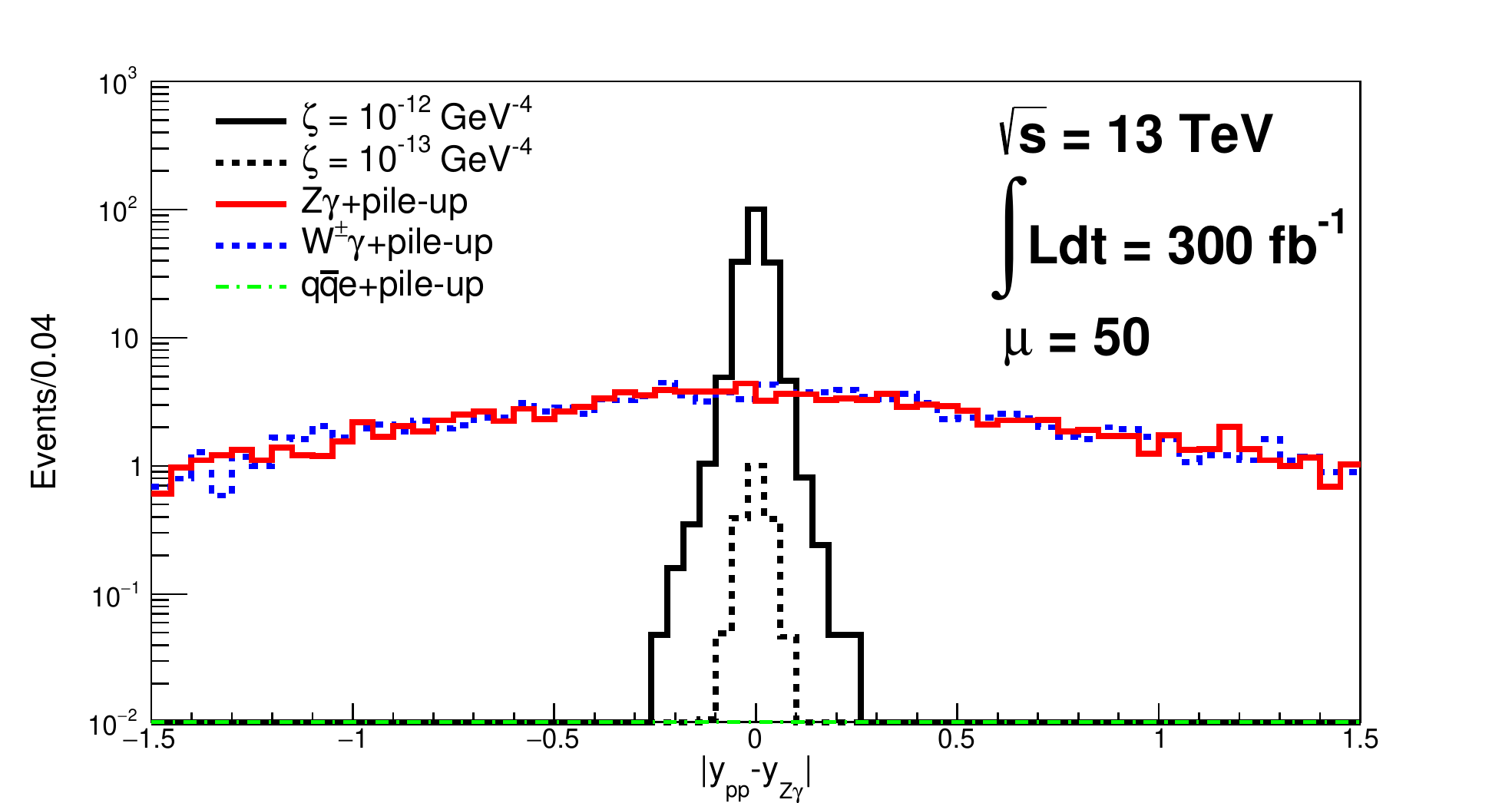}
\caption{\label{exclusivity_jets} Missing diproton mass $m_{pp} = \sqrt{\xi_1\xi_2 s}$ to central mass ratio distribution (Top) and rapidity difference distribution (Bottom) in the $jj\gamma$ channel for the signal and background within the acceptance $0.015<\xi_{1,2}<0.15$ considering two different coupling values after applying the requirement on $p_T$, invariant mass $m_{Z\gamma}$, $p_T$ ratios and angle separation according to Table \ref{qqbar_table}. The integrated luminosity is $300$ fb$^{-1}$  and the  average pile-up  is $\mu = 50$. The signal width is due to a combined effect of the reconstructed jet energy low resolution ($\approx$ 15\%) and the $\xi_{1,2}$ resolution from the proton detectors. The asymmetry on the $m_{pp}/m_{Z\gamma}$ distribution is due to the resolution on the jet energy.}
\end{figure}

\clearpage

Exploiting the full event kinematics with the forward proton detectors allows us to suppress the non-exclusive background in both channels with a high signal selection efficiency of $\sim 70\%$.  (for $\mu=50$ average pile-up interactions), as can be seen in Tables \ref{qqbar_table} and \ref{llbar_table} and Figures \ref{exclusivity_jets} and  \ref{exclusivity_leptons}. For a coupling value of $\zeta  = 4\cdot 10^{-13}$ GeV $^{-4}$, the signal cross section  within the proton taggers acceptance is $\sim 1.1$~fb. We expect about 25 events in $jj\gamma$ and $10$ events in $\ell\bar{\ell}\gamma$ channels for an integrated luminosity of 300 fb$^{-1}$ for this coupling value after selection cuts.

In addition, we include a similar study at a higher number of pile-up interactions per bunch crossing $\mu=200$ and an integrated luminosity of $3000\,\mathrm{fb}^{-1}$. We consider the $\ell\bar{\ell}\gamma$ final states in this scenario, since we do not obtain much improvement in the hadronic channel in comparison to the 300 fb$^{-1}$. We optimized the cuts for this case to increase the background rejection even further, as shown in Table \ref{llbar_table_highlumi}.

Let us note that a significant fraction of the $Z$ boson hadronic final states are reconstructed as a single jet, since the dijet system is boosted in the high mass regime where the signal is enhanced. The QCD background from  $q\gamma$ and $g\gamma$ final states is very large, and contributes to $\mathcal{O}(10^3)$ background events after selection cuts. For this reason, we restricted ourselves to the dijet final states in this study, 
We stress however that a more evolved jet substructure analysis can in principle efficiently discrimate between large-radius jets from a Z and from QCD \cite{Heinrich:2014kza} but this goes beyond our study.

\section{Expected sensitivities}
\label{se:sensitivities}

\begin{table}[t]
\centering
\begin{tabular}{|c||c|c||c|c|}
\hline
Coupling ($\mathrm{GeV}^{-4}$)             & \multicolumn{2}{c||}{$\zeta$ ($\tilde \zeta=0$)}    & \multicolumn{2}{c|}{$\zeta=\tilde \zeta$}         \\ \hline
Luminosity                   & \multicolumn{2}{c||}{300 fb$^{-1}$}                    & \multicolumn{2}{c|}{300 fb$^{-1}$}                   \\ \hline
Pile-up ($\mu$)              & \multicolumn{2}{c||}{50}                        & \multicolumn{2}{c|}{50}                       \\ \hline
Channels                     & 5 $\sigma$               & 95\% CL                & 5 $\sigma$               & 95\% CL               \\ \hline
$\ell\bar{\ell}\gamma$       & $2.8 \cdot 10^{-13}$ & $1.8\cdot 10^{-13}$     & $2.5\cdot 10^{-13}$  & $1.5\cdot 10^{-13}$  \\ \hline
$jj\gamma$             & $2.3 \cdot 10^{-13}$ & $1.5\cdot 10^{-13}$  & $2 \cdot 10^{-13}$    & $1.3 \cdot 10^{-13}$    \\ \hline
$jj\gamma \bigoplus \ell\bar{\ell}\gamma$                     & $1.93\cdot 10^{-13}$ & $1.2 \cdot 10^{-13}$ & $1.7 \cdot 10^{-13}$  & $1 \cdot 10^{-13}$  \\ \hline
\end{tabular}

\caption{ \label{reach_table_lowlumi} $5\sigma$ discovery and 95\% CL exclusion limits on $\zeta $, $\tilde\zeta $. Sensitivities are given for $300\, \mathrm{fb}^{-1}$ for $\mu=50$ pile-up interactions. The last row corresponds to the search of $Z\gamma$ production in both channels with their respective selection cuts.}
\end{table}

\begin{table}[t]
\centering
\begin{tabular}{|c||c|c||c|c|}
\hline
Coupling                      & \multicolumn{2}{c||}{$\zeta$ ( $\tilde \zeta = 0$ )}   & \multicolumn{2}{c|}{$\zeta=\tilde \zeta$}        \\ \hline
Luminosity                    & \multicolumn{2}{c||}{$3000\,\mathrm{fb}^{-1}$} & \multicolumn{2}{c|}{$3000\,\mathrm{fb}^{-1}$} \\ \hline
Pile-up                       & \multicolumn{2}{c||}{200}                      & \multicolumn{2}{c|}{200}                      \\ \hline
Channel                       & $5\sigma$             & 95 \% C.L.            & $5\sigma$             & 95\% C.L.             \\ \hline
$\ell\bar{\ell}\gamma$        & $1.8\cdot 10^{-13}$  & $1.1 \cdot 10^{-13}$    & $1.25 \cdot 10^{-13}$  & $7.8 \cdot 10^{-14}$    \\ \hline
\end{tabular}
\caption{\label{reach_table_highlumi} $5\sigma$ discovery and 95\% CL exclusion limits on $\zeta $, $\tilde\zeta $. Sensitivities are given for $3000\,\mathrm{fb}^{-1}$ and $\mu=200$ average pile-up interactions, which corresponds to the High Luminosity LHC. Sensitivities for the $jj\gamma$ channel are not quoted in this scenario, due to the high number of background events which compromises significantly the signal efficiency.}
\end{table}

\begin{figure}[t]
\centering
\includegraphics[width=0.7\textwidth]{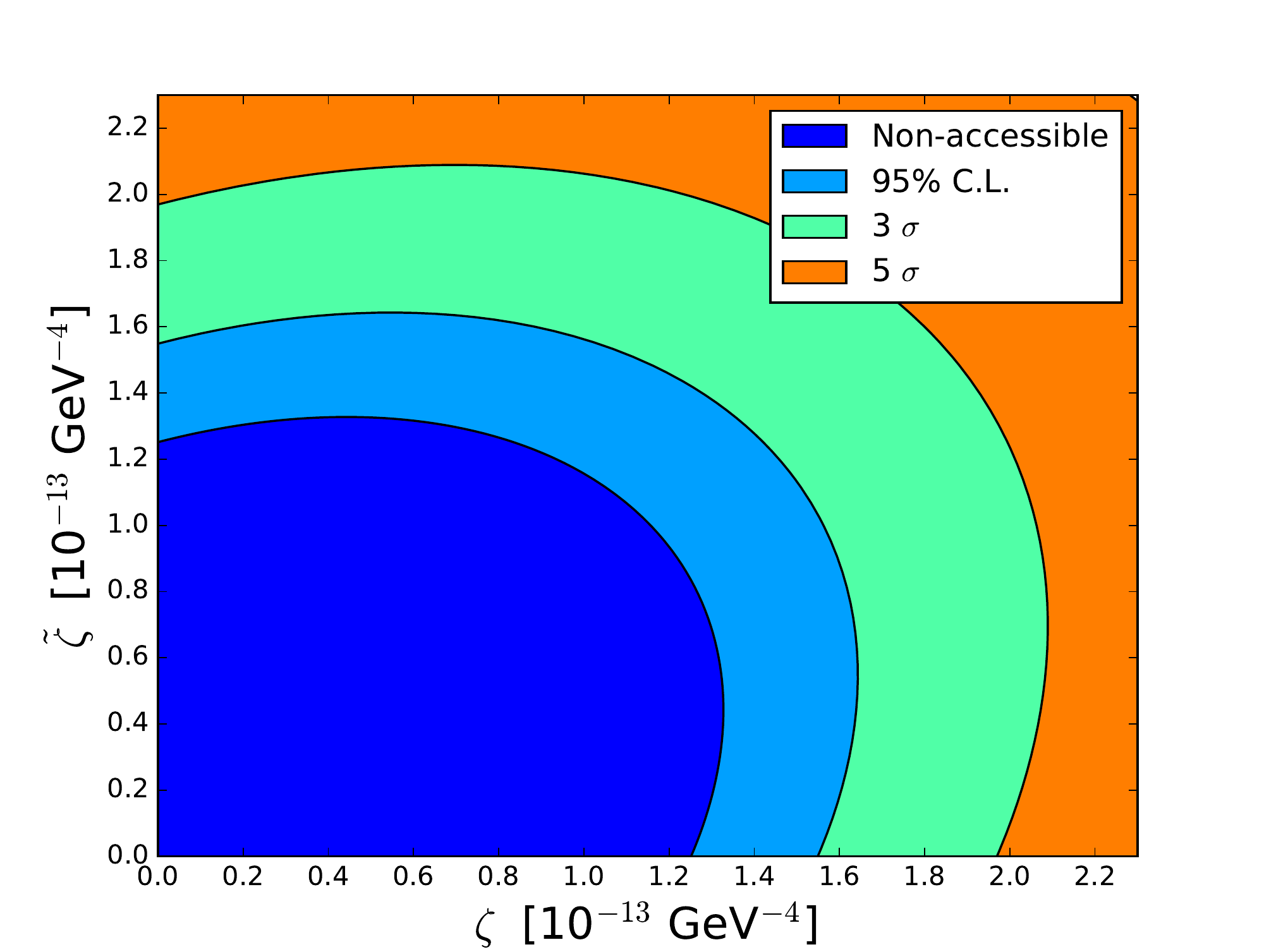}
\caption{ \label{zeta1_zeta3_plane} Sensitivity in the $\zeta -\tilde\zeta $ plane for 300 fb$^{-1}$ and $\mu = 50$. Orange, green and light blue can be probed at 5 $\sigma$, 3$\sigma$, and 95\% C.L. using proton tagging at the LHC. 
}
\end{figure}

Expected sensitivities to the $\zeta$, $ \tilde \zeta$ coefficients are shown in Tabs.~\ref{reach_table_lowlumi}, \ref{reach_table_highlumi} and in Fig.~\ref{zeta1_zeta3_plane}. 
The sensitivities are roughly of $\sim 2\cdot 10^{-13} $ GeV$^{-4}$ for both $300$~fb$^{-1}$ at low luminosity and $3000$~fb$^{-1}$ at high luminosity. The reach  at high luminosity is limited by the large pile-up.

The leptonic final state turns out to be the cleanest channel for this search with nearly background-free events without the use of timing detectors. 
Interestingly, it turns out that a good sensitivity is also obtained in the hadronic channel  at moderate pile-up, with only three background events after selection cuts. The remaining background events that may pass the selection cuts can be further rejected by a factor of $\sim$ 40 with the timing detectors to be in operation in AFP and CT-PPS, as discussed in Sec. \ref{considerations}.

The ${\cal B}(Z\rightarrow \gamma\gamma\gamma)$ branching ratio has been constrained at LEP \cite{Decamp:1991uy,Akrawy:1990zz, Abreu:1994du}. More recently a stronger bound  from ATLAS using 8~TeV data  has been  quoted in Ref.~\cite{Aad:2015bua}, ${\cal B}(Z\rightarrow \gamma\gamma\gamma)< 2.2 \cdot 10^{-6}$. This bound translates as a limit
\be
\sqrt{\zeta^2+\tilde\zeta^2 -\frac{\zeta \tilde \zeta}{2}  } < 1.3 \cdot 10^{-9} ~\textrm{GeV}^{-4}~~ (95\%{\rm CL}) \,.
 \label{eq:ATLAS_bound} 
\ee 
Imagining the same search is done at $13$ TeV data with $300$~fb$^{-1}$ in the same conditions, we expect very roughly an improvement by an order of magnitude of the bound of  Eq.~\ref{eq:ATLAS_bound}.  In addition, the current number of pile-up interactions at $13$~TeV sets a challenge to the measurement of $3\gamma$ final states. This remains far away from the expected sensitivities obtained in the exclusive channel at the same luminosity by roughly three orders of magnitudes.

Finally, let us compare the $\gamma Z$ channel sensitivity to exclusive $\gamma \gamma$ channel sensitivity estimated in Refs.~\cite{Fichet:2013gsa, Fichet2015}. We do so for the case of heavy neutral particles described in Sec.~\ref{se:NP}. 
Using the $\gamma Z$, $\gamma \gamma$ sensitivities at $300$ fb$^{-1}$ with no form factors, which are given respectively in Tab.~\ref{reach_table_lowlumi} of this paper and Tab.~3 of Ref.~\cite{Fichet2015}, we obtain that the neutral particle can be detected with 5$\sigma$ significance if
\bea
m<2.3\, {\rm TeV}\cdot \left(\frac{1\,\rm TeV}{\sqrt{f^{\gamma\gamma}f^{\gamma Z}}}\right) \, ~ ~ \textrm{in the} ~ \gamma Z ~ \textrm{ channel }\,, \label{eq:reach1}\\
m<4.5\, {\rm TeV}\cdot \left(\frac{1\,\rm TeV}{f^{\gamma\gamma}}\right)\, ~ ~ \textrm{in the} ~ \gamma \gamma ~ \textrm{ channel }\,.\label{eq:reach2}
\eea
Depending on the relative strength of the $f^{\gamma\gamma}$ and $f^{\gamma Z}$ couplings,  a similar reach can be obtained in the two channels.

We would like to stress that in order for these bounds to be the most sensitive probe of a new particle, one has to assume that the couplings to gluons are somewhat suppressed, as otherwise gluon fusion processes can provide stronger bounds. For instance, in the case of the KK graviton coupling universally to all SM fields, our 5$\sigma$ sensitivity in Eq.~(\ref{eq:reach2}) translates to $m_{\rm KK}<1.4 {\rm\ TeV} \sqrt{\kappa/0.1}$, where $\kappa$ is the universal KK graviton coupling strength. This bound is slightly weaker than standard searches. However, our method becomes more sensitive at large coupling, both because the resonance becomes too broad for standard searches, and also because we are sensitive to mass regions outside the kinematic reach of the LHC.
Finally, even if our method did not provide the primary discovery channel, 
Equations~(\ref{eq:reach1}) and (\ref{eq:reach2})  show that the $\gamma Z$ channel could efficiently determine whether
the underlying particle couples universally to $(B^{\mu\nu})^2$, $(W^{I,\mu\nu})^2$ (see discussion in Sec.~\ref{se:NP}).
Similar considerations apply to electroweak production modes. Because of gauge invariance, the $\gamma \gamma$ and $\gamma Z$ couplings 
also imply nonzero $WW$ and $ZZ$ coupings. Therefore, inelastic photon fusion as well as $WW$ and $ZZ$ fusion are always present. We have checked that of the latter, inelastic photon fusion has the largest cross section.~\footnote{For instance, in the case of the 750 GeV diphoton excess it was found that the inelastic photon fusion cross section is about 15-20 times the elastic one, while the correction to this from $ZZ$ and $WW$ contributions amounted to about 10\% \cite{Fichet:2016pvq}. Notice that the irreducible weak boson couplings have a different tensor structure than for instance in the case of the SM Higgs.} However, all of these production modes have to compete with a much larger background, which becomes particularly problematic the wider the resonance is.

The sensitivity to charged particles is fairly weak, unless large $d,Y$ or a large multiplicity are taken. 
For a vector in the $SU(2)_L$ adjoint for example, one has $d=3,Y=0$  and the reach on the mass is found to be $m \sim 120 $ GeV. For $N$ vectors, the bound would increase as $N^{1/4}$. 
On the other hand, this  bound on charged particles is very generic, as it depends only on the  quantum numbers of the underlying particle (see Eq.~\ref{eq:loop}). This measurement is thus quite complementary to direct searches for charged particles, which are very model-dependent.

\section{Conclusion} \label{se:conclusion}

The forward proton detectors recently installed at the LHC provide the opportunity of measuring the anomalous $\gamma\gamma\gamma Z$ coupling with unprecedented sensitivity, providing another high precision probe into the SM gauge sector.  We have estimated the discovery potential for exclusive $\gamma Z$ production at a center-of-mass energy of 13 TeV, for both low and high luminosity scenarios. 

We have computed the anomalous $\gamma \gamma \rightarrow \gamma Z$ rate induced by New Physics in the effective field theory framework. Contributions to the $\gamma\gamma\gamma Z$ couplings from tree-level exchange of neutral particles and from  loops of particles with arbitrary electroweak charge  have been  calculated.  
Prospects for both hadronic and leptonic channels have been evaluated. The tagging of the protons drastically reduces the background in all cases, and allows  studying the $Z$ decay in the hadronic channel in addition to the leptonic channel, which is usually  very challenging in the standard searches. 
For $300~$fb$^{-1}$ of integrated luminosity, similar sensitivities are found in both channels, and their combination amounts to  reach  the  anomalous couplings down to  $\zeta \sim 2\cdot 10^{-13}$~GeV$^{-4}$ with $5\sigma$ statistical significance.
This sensitivity goes beyond the one expected from the $Z\rightarrow \gamma\gamma\gamma$ decay searches at the LHC by roughly three orders of magnitude. 

The combined sensitivity in the $\gamma Z$ channel turns out to be roughly comparable to the sensitivity expected in the exclusive $\gamma \gamma$ channel. Combining the information from both channels would provide a powerful way to pin down New Physics scenarios. In particular, a number of New Physics candidates such as the radion or Kaluza Klein gravitons contribute to the $\gamma\gamma$ final state and not to the $\gamma Z$ one.

\section*{Acknowledgements} 

 We thank O. Kepka and M. Saimpert for helpful discussions on FPMC and pile-up simulation. CB thanks the financial support from the  grant of CR as a Foundation Distinguished Professor.  SF work was supported by the S\~ao Paulo Research Foundation (FAPESP) under grants \#2011/11973 and \#2014/21477-2.
\\
\\
\\

\bibliographystyle{JHEP} 
\bibliography{references.bib}

\providecommand{\href}[2]{#2}\begingroup\raggedright\begin{thebibliography}{10}

\bibitem{usww}
E.~Chapon, C.~Royon, and O.~Kepka, {\it {Anomalous quartic W W gamma gamma, Z Z
  gamma gamma, and trilinear WW gamma couplings in two-photon processes at high
  luminosity at the LHC}},  {\em Phys.Rev.} {\bf D81} (2010) 074003,
  [\href{http://arxiv.org/abs/0912.5161}{{\tt arXiv:0912.5161}}].

\bibitem{usw}
O.~Kepka and C.~Royon, {\it {Anomalous $W W \gamma$ coupling in photon-induced
  processes using forward detectors at the LHC}},  {\em Phys.Rev.} {\bf D78}
  (2008) 073005, [\href{http://arxiv.org/abs/0808.0322}{{\tt
  arXiv:0808.0322}}].

\bibitem{Gupta:2011be}
R.~S. Gupta, {\it {Probing Quartic Neutral Gauge Boson Couplings using
  diffractive photon fusion at the LHC}},  {\em Phys.Rev.} {\bf D85} (2012)
  014006, [\href{http://arxiv.org/abs/1111.3354}{{\tt arXiv:1111.3354}}].

\bibitem{Fichet:2013ola}
S.~Fichet and G.~von Gersdorff, {\it {Anomalous gauge couplings from composite
  Higgs and warped extra dimensions}},  {\em JHEP03(2014)102} (2013)
  [\href{http://arxiv.org/abs/1311.6815}{{\tt arXiv:1311.6815}}].

\bibitem{Senol:2013ym}
A.~Senol, {\it {$ZZ\gamma$ and $Z\gamma\gamma$ anomalous couplings in $\gamma
  p$ collision at the LHC}},  {\em Phys. Rev.} {\bf D87} (2013) 073003,
  [\href{http://arxiv.org/abs/1301.6914}{{\tt arXiv:1301.6914}}].

\bibitem{Fichet:2013gsa}
S.~Fichet, G.~von Gersdorff, O.~Kepka, B.~Lenzi, C.~Royon, and M.~Saimpert,
  {\it {Probing new physics in diphoton production with proton tagging at the
  Large Hadron Collider}},  {\em Phys. Rev.} {\bf D89} (2014) 114004,
  [\href{http://arxiv.org/abs/1312.5153}{{\tt arXiv:1312.5153}}].

\bibitem{Sun:2014qoa}
H.~Sun, {\it {Probe anomalous $tq\gamma$ couplings through single top
  photoproduction at the LHC}},  {\em Nucl.Phys.} {\bf B886} (2014) 691--711,
  [\href{http://arxiv.org/abs/1402.1817}{{\tt arXiv:1402.1817}}].

\bibitem{Sun:2014qba}
H.~Sun, {\it {Large Extra Dimension effects through Light-by-Light Scattering
  at the CERN LHC}},  {\em Eur.Phys.J.} {\bf C74} (2014) 2977,
  [\href{http://arxiv.org/abs/1406.3897}{{\tt arXiv:1406.3897}}].

\bibitem{Sun:2014ppa}
H.~Sun, {\it {Dark Matter Searches in Jet plus Missing Energy in $\rm \gamma p$
  collision at CERN LHC}},  {\em Phys.Rev.} {\bf D90} (2014) 035018,
  [\href{http://arxiv.org/abs/1407.5356}{{\tt arXiv:1407.5356}}].

\bibitem{Senol:2013lca}
A.~Senol, {\it {Anomalous quartic $WW\gamma\gamma$ and $ZZ\gamma\gamma$
  couplings in $\gamma p$ collision at the LHC}},  {\em Int. J. Mod. Phys.}
  {\bf A29} (2014), no.~26 1450148, [\href{http://arxiv.org/abs/1311.1370}{{\tt
  arXiv:1311.1370}}].

\bibitem{Sahin:2014dua}
I.~Sahin, M.~Koksal, S.~C. Inan, A.~A. Billur, B.~Sahin, P.~Tektas, E.~Alici,
  and R.~Yildirim, {\it {Graviton production through photon-quark scattering at
  the LHC}},  {\em Phys. Rev.} {\bf D91} (2015) 035017,
  [\href{http://arxiv.org/abs/1409.1796}{{\tt arXiv:1409.1796}}].

\bibitem{Inan:2014mua}
S.~C. Inan, {\it {Dimension-six anomalous $tq\gamma$ couplings in
  $\gamma\gamma$ collision at the LHC}},  {\em Nucl. Phys.} {\bf B897} (2015)
  289--301, [\href{http://arxiv.org/abs/1410.3609}{{\tt arXiv:1410.3609}}].

\bibitem{Fichet:2015nia}
S.~Fichet, {\it {Prospects for new physics searches at the LHC in the forward
  proton mode}},  {\em Acta Phys. Polon. Supp.} {\bf 8} (2015) 811,
  [\href{http://arxiv.org/abs/1510.01004}{{\tt arXiv:1510.01004}}].

\bibitem{Fichet2015}
S.~Fichet, G.~von Gersdorff, B.~Lenzi, C.~Royon, and M.~Saimpert, {\it
  {Light-by-light scattering with intact protons at the LHC: from Standard
  Model to New Physics}},  {\em JHEP} {\bf 02} (2015) 165,
  [\href{http://arxiv.org/abs/1411.6629}{{\tt arXiv:1411.6629}}].

\bibitem{Senol:2014vta}
A.~Senol and M.~Köksal, {\it {Analysis of anomalous quartic $WWZ\gamma$
  couplings in $\gamma p$ collision at the LHC}},  {\em Phys. Lett.} {\bf B742}
  (2015) 143--148, [\href{http://arxiv.org/abs/1410.3648}{{\tt
  arXiv:1410.3648}}].

\bibitem{Fichet:2016clq}
S.~Fichet, {\it {Shining Light on Polarizable Dark Particles}},
  \href{http://arxiv.org/abs/1609.01762}{{\tt arXiv:1609.01762}}.

\bibitem{Fichet:2016pvq}
S.~Fichet, G.~von Gersdorff, and C.~Royon, {\it {Measuring the Diphoton
  Coupling of a 750 GeV Resonance}},  {\em Phys. Rev. Lett.} {\bf 116} (2016),
  no.~23 231801, [\href{http://arxiv.org/abs/1601.01712}{{\tt
  arXiv:1601.01712}}].

\bibitem{Esmaili:2016enf}
A.~Esmaili, S.~Khatibi, and M.~Mohammadi~Najafabadi, {\it {Constraining the
  monochromatic gamma-rays from dark matter annihilation by the LHC}},
  \href{http://arxiv.org/abs/1611.09320}{{\tt arXiv:1611.09320}}.

\bibitem{Khoze:2017igg}
V.~A. Khoze, A.~D. Martin, and M.~G. Ryskin, {\it {Can invisible objects be
  `seen' via forward proton detectors at the LHC?}},
  \href{http://arxiv.org/abs/1702.05023}{{\tt arXiv:1702.05023}}.

\bibitem{ATLAS_detector_paper}
{\bf ATLAS} Collaboration, G.~Aad et~al., {\it {The ATLAS Experiment at the
  CERN Large Hadron Collider}},  {\em JINST} {\bf 3} (2008) S08003.

\bibitem{Royon:2015tfa}
C.~Royon and N.~Cartiglia, {\it {The AFP and CT-PPS projects}},  {\em Int. J.
  Mod. Phys. A29} {\bf A29} (2014), no.~28 1446017,
  [\href{http://arxiv.org/abs/1503.04632}{{\tt arXiv:1503.04632}}].

\bibitem{Jikia:1993pg}
G.~Jikia and A.~Tkabladze, {\it {gamma Z pair production at the photon linear
  collider}},  {\em Phys. Lett.} {\bf B332} (1994) 441--447,
  [\href{http://arxiv.org/abs/hep-ph/9312274}{{\tt hep-ph/9312274}}].

\bibitem{Laursen:1980ba}
M.~L. Laursen, K.~O. Mikaelian, and M.~A. Samuel, {\it {Z0 decay into three
  gluons}},  {\em Phys. Rev.} {\bf D23} (1981) 2795.

\bibitem{vanderBij:1988ac}
J.~J. van~der Bij and E.~W.~N. Glover, {\it {$Z$ Boson Production and Decay via
  Gluons}},  {\em Nucl. Phys.} {\bf B313} (1989) 237--257.

\bibitem{Baillargeon:1991ia}
M.~Baillargeon and F.~Boudjema, {\it {Contribution of the bosonic loops to the
  three photon decay of the Z}},  {\em Phys. Lett.} {\bf B272} (1991) 158--162.

\bibitem{Belanger:1999aw}
G.~Belanger, F.~Boudjema, Y.~Kurihara, D.~Perret-Gallix, and A.~Semenov, {\it
  {Bosonic quartic couplings at LEP-2}},  {\em Eur. Phys. J.} {\bf C13} (2000)
  283--293, [\href{http://arxiv.org/abs/hep-ph/9908254}{{\tt hep-ph/9908254}}].

\bibitem{Eboli:2006wa}
O.~J.~P. Eboli, M.~C. Gonzalez-Garcia, and J.~K. Mizukoshi, {\it {p p ---> j j
  e+- mu+- nu nu and j j e+- mu-+ nu nu at O( alpha(em)**6) and O(alpha(em)**4
  alpha(s)**2) for the study of the quartic electroweak gauge boson vertex at
  CERN LHC}},  {\em Phys. Rev.} {\bf D74} (2006) 073005,
  [\href{http://arxiv.org/abs/hep-ph/0606118}{{\tt hep-ph/0606118}}].

\bibitem{Adams:2006sv}
A.~Adams, N.~Arkani-Hamed, S.~Dubovsky, A.~Nicolis, and R.~Rattazzi, {\it
  {Causality, analyticity and an IR obstruction to UV completion}},  {\em JHEP}
  {\bf 10} (2006) 014, [\href{http://arxiv.org/abs/hep-th/0602178}{{\tt
  hep-th/0602178}}].

\bibitem{Baillargeon:1995dg}
M.~Baillargeon, F.~Boudjema, E.~Chopin, and V.~Lafage, {\it {New physics with
  three photon events at LEP}},  {\em Z. Phys.} {\bf C71} (1996) 431--442,
  [\href{http://arxiv.org/abs/hep-ph/9506396}{{\tt hep-ph/9506396}}].

\bibitem{Agashe:2014kda}
{\bf Particle Data Group} Collaboration, K.~Olive et~al., {\it {Review of
  Particle Physics}},  {\em Chin.Phys.} {\bf C38} (2014) 090001.

\bibitem{Terazawa:1973tb}
H.~Terazawa, {\it {Two photon processes for particle production at
  high-energies}},  {\em Rev.Mod.Phys.} {\bf 45} (1973) 615--662.

\bibitem{Budnev}
V.~Budnev, I.~Ginzburg, G.~Meledin, and V.~Serbo, {\it {The Two photon particle
  production mechanism. Physical problems. Applications. Equivalent photon
  approximation}},  {\em Phys.Rept.} {\bf 15} (1975) 181--281.

\bibitem{survival1}
V.~Khoze, A.~Martin, and M.~Ryskin, {\it {Prospects for new physics
  observations in diffractive processes at the LHC and Tevatron}},  {\em
  Eur.Phys.J.} {\bf C23} (2002) 311--327,
  [\href{http://arxiv.org/abs/hep-ph/0111078}{{\tt hep-ph/0111078}}].

\bibitem{survival2}
E.~Gotsman, E.~Levin, U.~Maor, E.~Naftali, and A.~Prygarin, {\it {Survival
  probability of large rapidity gaps}},
  \href{http://arxiv.org/abs/hep-ph/0511060}{{\tt hep-ph/0511060}}.

\bibitem{cms}
{CMS and TOTEM collaboration, CERN-LHCC-2014-021}, {\it {CMS-TOTEM Precision
  Proton Spectrometer}}, .

\bibitem{Timing}
C.~Royon, M.~Saimpert, O.~Kepka, and R.~Zlebcik, {\it {Timing detectors for
  proton tagging at the LHC}},  {\em {Acta Physica Polonica B Proceedings
  supplement}} {\bf 7} (2014) 735.

\bibitem{Khachatryan:2015iwa}
{\bf CMS} Collaboration, V.~Khachatryan et~al., {\it {Performance of Photon
  Reconstruction and Identification with the CMS Detector in Proton-Proton
  Collisions at sqrt(s) = 8 TeV}},  {\em JINST} {\bf 10} (2015), no.~08 P08010,
  [\href{http://arxiv.org/abs/1502.02702}{{\tt arXiv:1502.02702}}].

\bibitem{Barter:2016oan}
W.~Barter, {\it {Inclusive single gauge boson production in $\sqrt{s} = 7, 8$
  and $13$ TeV proton-proton collisions}},  in {\em {51st Rencontres de Moriond
  on EW Interactions and Unified Theories La Thuile, Italy, March 12-19,
  2016}}, 2016.
\newblock \href{http://arxiv.org/abs/1605.02487}{{\tt arXiv:1605.02487}}.

\bibitem{Schroeder:2015ila}
{\bf CMS} Collaboration, M.~Schroeder, {\it {Performance of jets at CMS}},
  {\em J. Phys. Conf. Ser.} {\bf 587} (2015), no.~1 012004.

\bibitem{Berta:2016ukt}
{\bf ATLAS} Collaboration, P.~Berta, {\it {ATLAS jet and missing-ET
  reconstruction, calibration, and performance}},  {\em Nucl. Part. Phys.
  Proc.} {\bf 273-275} (2016) 1121--1126.

\bibitem{FPMC}
M.~Boonekamp, A.~Dechambre, V.~Juranek, O.~Kepka, M.~Rangel, et~al., {\it
  {FPMC: A Generator for forward physics}},
  \href{http://arxiv.org/abs/1102.2531}{{\tt arXiv:1102.2531}}.

\bibitem{Sjostrand:2014zea}
T.~Sjöstrand, S.~Ask, J.~R. Christiansen, R.~Corke, N.~Desai, P.~Ilten,
  S.~Mrenna, S.~Prestel, C.~O. Rasmussen, and P.~Z. Skands, {\it {An
  Introduction to PYTHIA 8.2}},  {\em Comput. Phys. Commun.} {\bf 191} (2015)
  159--177, [\href{http://arxiv.org/abs/1410.3012}{{\tt arXiv:1410.3012}}].

\bibitem{Fastjet}
M.~Cacciari, G.~P. Salam, and G.~Soyez, {\it {FastJet User Manual}},  {\em Eur.
  Phys. J.} {\bf C72} (2012) 1896, [\href{http://arxiv.org/abs/1111.6097}{{\tt
  arXiv:1111.6097}}].

\bibitem{Kepka_pileup}
O.~Kepka, {\it {Impact of multiple pp interactions on proton tagging,
  presentation at the LHC Working Group on Forward Physics and Diffraction}},
  {\em May 15-16 2013}.

\bibitem{Khoze:2001xm}
V.~Khoze, A.~Martin, and M.~Ryskin, {\it {Prospects for new physics
  observations in diffractive processes at the LHC and Tevatron}},  {\em
  Eur.Phys.J.} {\bf C23} (2002) 311--327,
  [\href{http://arxiv.org/abs/hep-ph/0111078}{{\tt hep-ph/0111078}}].

\bibitem{Jikia:1997yt}
G.~Jikia, {\it {Electroweak gauge boson production at gamma gamma collider}},
  {\em Turk. J. Phys.} {\bf 22} (1998) 705--713,
  [\href{http://arxiv.org/abs/hep-ph/9710459}{{\tt hep-ph/9710459}}].

\bibitem{Heinrich:2014kza}
J.~J. Heinrich, {\it {Reconstruction of boosted $W^{\pm}$ and $Z^{0}$ bosons
  from fat jets}},  Master's thesis, Bohr Inst., 2014-08-15.

\bibitem{Decamp:1991uy}
{\bf ALEPH} Collaboration, D.~Decamp et~al., {\it {Searches for new particles
  in $Z$ decays using the ALEPH detector}},  {\em Phys. Rept.} {\bf 216} (1992)
  253--340.

\bibitem{Akrawy:1990zz}
{\bf OPAL} Collaboration, M.~Z. Akwawy et~al., {\it {Measurement of the
  cross-sections of the reactions $e^+ e^- \rightarrow \gamma \gamma$ and $e^+
  e^- \rightarrow \gamma \gamma \gamma$ at LEP}},  {\em Phys. Lett.} {\bf B257}
  (1991) 531--540.

\bibitem{Abreu:1994du}
{\bf DELPHI} Collaboration, P.~Abreu et~al., {\it {Measurement of the $e^+ e^-
  \rightarrow \gamma \gamma (\gamma)$ cross-section at LEP energies}},  {\em
  Phys. Lett.} {\bf B327} (1994) 386--396.

\bibitem{Aad:2015bua}
{\bf ATLAS} Collaboration, G.~Aad et~al., {\it {Search for new phenomena in
  events with at least three photons collected in $pp$ collisions at $\sqrt{s}$
  = 8 TeV with the ATLAS detector}},  {\em Eur. Phys. J.} {\bf C76} (2016),
  no.~4 210, [\href{http://arxiv.org/abs/1509.05051}{{\tt arXiv:1509.05051}}].

\end{thebibliography}\endgroup

\end{document}